\documentclass[entropy,review,accept,moreauthors,12pt,a4paper]{mdpi}
\lastpage{x}
\doinum{10.3390/------}
\pubvolume{xx}
\pubyear{2014}
\history{Received: xx / Accepted: xx / Published: xx}
\usepackage{graphicx}
\usepackage{subfigure}

\Title{FACTORIZATION AND CRITICALITY IN THE ANISOTROPIC XY CHAIN VIA CORRELATIONS}

\Author{Bar{\i}\c{s} \c{C}akmak $^{1,}$*, G\"{o}ktu\u{g} Karpat $^{1}$ and Felipe F. Fanchini $^{1}$}

\address{%
$^{1}$ Faculdade de Ci\^encias, UNESP - Universidade Estadual Paulista, Bauru, SP, 17033-360, Brazil}

\corres{cakmakb@sabanciuniv.edu}

\abstract{In this review, we discuss the zero and finite temperature behavior of various bipartite quantum and total correlation measures, the skew information-based quantum coherence, and the local quantum uncertainty in the thermal ground state of the one-dimensional anisotropic XY model in transverse magnetic field. We compare the ability of considered measures to correctly detect or estimate the quantum critical point and the non-trivial factorization point possessed by the spin chain.}


\begin{document}

%%%%%%%%%%%%%%%%%%%%%%%%%%%%%%%%%%%%%%%%%%

\section{Introduction}

In recent years, methods of quantum information science have been extensively used in numerous fields of physics, especially in condensed matter theory \cite{amico08}. One of the most important tools borrowed from quantum information theory is based on the concept of entanglement, which is thought to be the characteristic trait of quantum mechanics. Entanglement has been investigated in many physical settings, and it has been widely considered to be the main resource in most of the quantum information processing tasks~\cite{concur1,concur2,entreview}. However, in the last decade, it has been shown that entanglement is not the only meaningful correlation present in quantum states, that is separable quantum states can also be exploited to provide a quantum mechanical speed-up over classical methods. Quantum discord (QD) is the most significant correlation measure that can capture quantum correlations more general than entanglement~\cite{discordreview,discord1,discord2}. The demonstration of the usefulness of QD as a novel quantum resource has triggered a new line of research, which aims to further characterize the quantum correlations beyond entanglement in the quantum information community. Following QD, many other quantum and total correlation measures have been introduced and widely studied from several perspectives \cite{discord3,luo11,WYSIcorr,discord4}.

Quantum phase transitions (QPT) are sudden changes occurring in ground states of many-body quantum systems, which are driven by a control parameter in the Hamiltonian describing the system~\cite{qpt}. The key feature of QPTs is that they have their roots in purely quantum fluctuations due to the Heisenberg uncertainty relation, in contrast to classical phase transitions, which are only driven by thermal fluctuations. In principle, QPTs occur at absolute zero temperature, which is unattainable experimentally in practice. However, they are more than a mere theoretical construct, since the signatures of QPTs can also be observed at finite and experimentally accessible temperatures. In fact, traces of QPTs are detectable when the thermal fluctuations are smaller than the de Broglie wavelength of the particles in the system. QPTs are related to the energy level crossings occurring in the ground state of the many-body system, which result in the discontinuities in the ground state energy. The order of the phase transition is then determined based on this discontinuity. While a discontinuity in the ground state energy signals a first-order transition, a discontinuity in the first derivative of the ground state energy can be observed when the second-order phase transition takes place.

Since quantum spin chains possess several different kinds of QPTs, they are natural candidates for studying quantum critical phenomena in many-body systems. On the other hand, another peculiar property of spin chains in a transverse magnetic field is the factorization phenomena \cite{factor1}. Although the ground states of such systems are in general entangled, for some specific Hamiltonian parameters, the ground state becomes completely factorized. This phenomenon has its roots in the symmetry of the ground state of the system under consideration. The detection and understanding of this phenomena are rather non-trivial and have been the subject of many works \cite{factor2,factor3,factor4,factor5,factorcon1,factorcon2,factorcon3,factorcon4,factorcon5}.

In this review, we intend to cover the literature on the behavior of correlations in various quantum spin chains, both near QPT and the factorization point (FP). We summarize the results on the key contributions in the field in detail and discuss their implications both at zero~ \cite{et01,et02,wu,et03,et04,et05,et06,et07,et08,et09,et010,et011,et012,et013,et014,et015,et016,et017,et018,et019,et020,et021,et022,dt01,dt02,dt03,dt04,dt05,dt06,
dt07,dt08,dt09,dt010,dt011,cheng,dt012,dt013} and finite temperatures~ \cite{werlang1,werlang2,werlang3,nott01,nott02,nott03,nott04,nott05,nott06,ssbandnott01,ssbandnott02}. The remainder of the review focuses on some specific results concerning the correlations, coherence and uncertainty in the 1D anisotropic XY model in the transverse field. In particular, we summarize the analysis of the behavior of four quantum and total correlation measures at zero and finite temperatures~\cite{nott03,nott06}. In order to quantify total correlations, a general measure of non-locality \cite{luo11} and a measure based on Wigner--Yanase skew information \cite{WYSIcorr} are used. Quantum correlations are quantified by an experimentally lower bound of the geometric version of QD \cite{discord4}, and entanglement is quantified by concurrence \cite{concur1,concur2}. On the other hand, coherence and uncertainty are quantified by the regular and an optimized version of Wigner--Yanase skew information, respectively~ \cite{WYSIqc,WYSIcoh}.

This review is organized as follows. In Section 2, the spin-1/2 anisotropic XY model in a transverse magnetic field is introduced. QPTs, factorization phenomena and the symmetries of the ground state are discussed. Section 3 introduces the quantum and total correlation measures used in this review together with their physical interpretations. Section 4 begins with a comprehensive review of the works on quantum correlations in quantum spin chains and continues with some detailed specific results on the XY model in a transverse magnetic field. Section 5 includes our conclusion.

%%%%%%%%%%%%%%%%%%%%%%%%%%%%%%%%%%%%%%%%%%

\section{Spin-1/2 Anisotropic XY Chain in a Transverse Field}

The Hamiltonian of the one-dimensional anisotropic spin-1/2 XY chain in a transverse magnetic field is given by:
\begin{equation} \nonumber
 H=-\frac{\lambda}{2}\sum_{j=1}^{N} [(1+\gamma)\sigma^j_x \sigma^{j+1}_x +(1-\gamma)\sigma^{j}_{y}\sigma^{j+1}_{y}]-\sum_{j=1}^{N}\sigma^{j}_{z}
\end{equation}
where $\sigma^j_{x,y,z}$ are the usual Pauli operators at the $j$-th site, $\lambda$ denotes the strength of the inverse magnetic field, $\gamma \in [0,1]$ is the anisotropy parameter and $N$ is the number of spins. The Hamiltonian, which is in the Ising universality class for $\gamma>0$, reduces to the XX chain for $\gamma=0$ and to the Ising model in transverse field for $\gamma=1$. The system possesses a second order QPT in the parameter interval $0<\gamma\leq 1$ at the critical external field $\lambda_c=1$ that separates a paramagnetic (disordered) phase from a ferromagnetic (ordered) phase. There is another QPT in the region $\lambda>1$ at $\gamma_c=0$, which is of the Berezinskii--Kosterlitz--Thouless type and separates a ferromagnet ordered phase in the x-direction from a ferromagnet ordered phase in the y-direction. However, this transition will not be deeply explored in this review.

Although the ground state of the considered model is entangled in general, there exists a non-trivial factorization line corresponding to $\gamma^2+\lambda^{-2}=1$. Therefore, the ground state becomes completely factorized at any field point satisfying the equation,
\begin{equation} \label{factor}
\lambda_f=\frac{1}{\sqrt{1-\gamma^2}}.
\end{equation}

The diagonalization procedure of the XY model includes the well-established techniques of Jordan--Wigner and Bogoliubov transformations \cite{XYsol1,XYsol2}. The model Hamiltonian is invariant under a $\pi$ rotation around the z-axis of all spins (a parity transformation), which can also be stated by saying that the system exhibits global $Z_2$ symmetry. Taking this symmetry into account together with the translational invariance of the system, the reduced density matrix of two spins have an X-shaped form, and in terms of the magnetization and two-spin correlation functions, it can be written as:
%\begin{equation} \label{twospindm}
% \rho_{0r} = \frac{1}{4}[I+\langle\sigma_z\rangle (\sigma^0_z+\sigma^r_z)]+\frac{1}{4}\sum_{\alpha=x,y,z}\langle\sigma^0_{\alpha}\sigma^r_{\alpha}\rangle
% \sigma^0_{\alpha}\sigma^r_{\alpha},
%\end{equation}
%where $I$ is the four-dimensional identity matrix. The explicit form of the density matrix is known as X-shaped and can be written as
\begin{equation} \label{twospindm}
\begin{split}
 \rho^{0r} &= \begin{pmatrix}
 \rho_{11} & 0 & 0 & \rho_{14} \\
 0 & \rho_{22} & \rho_{23} & 0 \\
 0 & \rho_{23} & \rho_{22} & 0 \\
 \rho_{14} & 0 & 0 & \rho_{44} \end{pmatrix}\\ &= \begin{pmatrix}
 \frac{1}{4}+\frac{\langle\sigma^z\rangle}{2}+\frac{\langle\sigma_0^z\sigma_r^z\rangle}{4} & 0 & 0 & \frac{\langle\sigma_0^x\sigma_r^x\rangle-\langle\sigma_0^y\sigma_r^y\rangle}{4} \\
 0 & \frac{1-\langle\sigma_0^z\sigma_r^z\rangle}{4} & \frac{\langle\sigma_0^x\sigma_r^x\rangle+\langle\sigma_0^y\sigma_r^y\rangle}{4} & 0 \\
 0 & \frac{\langle\sigma_0^x\sigma_r^x\rangle+\langle\sigma_0^y\sigma_r^y\rangle}{4} & \frac{1-\langle\sigma_0^z\sigma_r^z\rangle}{4} & 0 \\
 \frac{\langle\sigma_0^x\sigma_r^x\rangle-\langle\sigma_0^y\sigma_r^y\rangle}{4} & 0 & 0 & \frac{1}{4}-\frac{\langle\sigma^z\rangle}{2}+\frac{\langle\sigma_0^z\sigma_r^z\rangle}{4} \end{pmatrix}.
\end{split}
\end{equation}
Here, $r$ denotes the distance between the spins, and the two-spin reduced density matrix is only dependent on the distance between the spins, $r=|i-j|$, due to the translational invariance. The magnetization and two-spin correlation functions are given as \cite{XYsol1,XYsol2}:
\begin{equation} \nonumber
\langle\sigma^z\rangle = -\int_0^{\pi} \frac{(1+\lambda\cos \phi)\tanh(\beta\omega_{\phi})}{2\pi\omega_{\phi}}d\phi,
\end{equation}
\begin{align} \nonumber
\langle\sigma_0^x\sigma_r^x\rangle = & \begin{vmatrix}
G_{-1} & G_{-2} & \cdots & G_{-r} \\
G_0 & G_{-1} & \cdots & G_{-r+1} \\
\vdots & \vdots & \ddots & \vdots \\
G_{r-2} & G_{r-3} & \cdots & G_{-1} \end{vmatrix}, \\ \nonumber
\langle\sigma_0^y\sigma_r^y\rangle = & \begin{vmatrix}
G_1 & G_0 & \cdots & G_{-r+2} \\
G_2 & G_1 & \cdots & G_{-r+3} \\
\vdots & \vdots & \ddots & \vdots \\
G_r & G_{r-1} & \cdots & G_1 \end{vmatrix}, \\ \nonumber
\langle\sigma_0^z\sigma_r^z\rangle = & \langle\sigma^z\rangle ^2-G_rG_{-r},
\end{align}
where the function $G_r$ is given as follows:
\begin{align} \nonumber
 G_r= &\int_0^{\pi} \frac{\tanh(\beta\omega_{\phi})\cos(r\phi)(1+\lambda\cos \phi)}{2\pi\omega_{\phi}}d\phi \\ \nonumber
  &-\gamma\lambda\int_0^{\pi} \frac{\tanh(\beta\omega_{\phi})\sin(r\phi)\sin(\phi)}{2\pi\omega_{\phi}}d\phi,
\end{align}
and $\omega_{\phi}=\sqrt{(\gamma\lambda\sin\phi)^2+(1+\lambda\cos \phi)^2}/2$ with $\beta =1/kT$ being the inverse temperature. Tracing out one of the spins, we can write the single-spin reduced density matrix as follows:
\begin{align} \label{singlespindm}
\rho_0=\rho_i=\frac{1}{2}\begin{pmatrix} 1+ \langle\sigma^z\rangle & 0\\ 0 & 1- \langle\sigma^z\rangle \end{pmatrix},
\end{align}
where $\langle\sigma^z\rangle$ is the transverse magnetization, and the density matrix is written on the basis of the eigenvectors of $\sigma_z$. Note that due to the translational invariance of our system, it is not important which spin we trace out. Indeed, all of the single-spin reduced density matrices are the same in this case.

The paramagnet-ferromagnet transition at $\lambda_c=1$ in the $0<\gamma\leq 1$ region has important consequences regarding the ground state of the system. At the CP%please define
 of the QPT, $Z_2$ symmetry is broken with the expectation value of the magnetization in the x-direction, $\langle S_x\rangle$, being the order parameter for this transition with a non-zero value for $\lambda>1$, and the ground state of the system becomes two-fold degenerate. In real physical settings, the system chooses one of the degenerate ground states as the real ground state due to some small external perturbation, which is called the mechanism of spontaneous symmetry breaking (SSB). However, in this case, the ground state of the system does not possess the symmetries of the Hamiltonian, for example the $Z_2$ symmetry in our case, and the reduced density matrix given in Equation (\ref{twospindm}) no longer has the X-shaped form. In the vast majority of the studies present in the literature, the effect of the broken symmetry is not taken into account, apart from a few exceptions \cite{ssbandnott01,ssbandnott02,ssb1,ssb2,ssb3,ssb4}, due to the complications it introduces into the calculations. We should note that, throughout this paper, we also neglect the effects of SSB and consider the so-called thermal ground state of the system. The thermal ground state is an equal mixture of these two degenerate states. Indeed, it is nothing but the limit $\beta\rightarrow\infty$ of the canonical ensemble,
\begin{equation} \label{thermal}
\rho= \lim_{\beta\rightarrow\infty} \frac{e^{-\beta H}}{Z},
\end{equation}
where $Z$ is the partition function. It is important to note that the thermal ground state is actually the same as the real ground state when there is no ground state degeneracy. Therefore, in the disordered phase, we are working with the real ground state of the system.

\section{Correlations, Coherence and Uncertainty}

In this section, we introduce the measures of correlations, coherence and uncertainty, which are central to this work. In the following sections, we will extensively investigate the behavior of the measures introduced here in the context of criticality and factorization in the anisotropic XY model.

\subsection{Geometric Measure of Quantum Discord}

Quantum discord is the first and the most popular quantum correlation measure in the zoo of measures that are more general than entanglement \cite{discord1,discord2}. It has been defined in terms of the discrepancy between the two classically equivalent generalizations of the quantum mutual information and can be present in states that contain no entanglement.

The amount of total correlations in a given quantum system can be quantified by the mutual information:
\begin{equation} \label{mi}
I(\rho^{AB})=S(\rho^A)+S(\rho^B)-S(\rho^{AB}),
\end{equation}
where $\rho^{AB}$ is the density matrix of the total system, $\rho^a$ and $\rho^b$ are the reduced density matrices of the subsystems and $S(\rho)=-\text{tr}\rho\log_2\rho$ is the von Neumann entropy. On the other hand, it is possible to quantify the classical correlations contained in a quantum system as follows \cite{discord2}:
\begin{equation}
C(\rho^{ab})=S(\rho^a)-\underset{\{\Pi_k^b\}}{\min}\sum_kp_kS(\rho_k^a),
\end{equation}
where \{$\Pi_k^b$\} is the set of most general quantum measurements performed on subsystem $b$ and $\rho_k^a=(I\otimes \Pi_k^b)\rho^{ab}(I\otimes \Pi_k^b)/p_k$ are the post-measurement states of subsystem $a$ after obtaining the outcome $k$ with probability $p_k=\text{tr}(I^a\otimes\Pi_k^b \rho^{ab})$ from the measurements made on subsystem $b$. Quantum discord is defined as the difference between the total and classical correlations, which, as a result, quantifies the quantum correlations:
\begin{equation}
\delta(\rho^{ab})=I(\rho^{ab})-C(\rho^{ab}).
\end{equation}

However, calculation of quantum discord is a rather difficult task due to the complex optimization procedure in the evaluation of the classical correlations, $C$. In order to overcome this difficulty, a geometric measure of quantum discord (GMQD) has been introduced \cite{discord3}, which measures the distance between the given state and the set of zero-discord states. Mathematically, it can be expressed as follows:
\begin{equation}
D_{G}(\rho^{ab})=2\min_{\chi}\|\rho^{ab}-\chi\|^{2},
\end{equation}
where the minimum is taken over the set of zero-discord states, $\chi$, and $\|.\|^{2}$ denotes the square of the Hilbert--Schmidt norm.

The representation of a general bipartite state in a Bloch basis can always be written as:
\begin{align}
\rho^{ab} &= \frac{1}{\sqrt{mn}} \frac{I^{a}}{\sqrt{m}} \otimes \frac{I^{b}}{\sqrt{n}}+ \sum_{i=1}^{m^2-1}x_{i}X_{i} \otimes \frac{I^{b}}{\sqrt{n}} \nonumber \\
 &+\frac{I^{a}}{\sqrt{m}} \otimes \sum_{j=1}^{n^2-1}y_{j}Y_{j} + \sum_{i=1}^{m^2-1}\sum_{j=1}^{n^2-1} t_{ij} X_{i} \otimes Y_{j},
\end{align}
where the matrices $\{X_{i}: i=0,1,\cdots,m^{2}-1\}$ and $\{Y_{j}: j=0,1,\cdots,n^{2}-1\}$, satisfying $\textmd{tr}(X_{k}X_{l})=\textmd{tr}(Y_{k}Y_{l})=\delta_{kl}$, define an orthonormal Hermitian operator basis associated with the subsystems $a$ and $b$, respectively. The components of the local Bloch vectors $\vec{x}=\{x_{i}\}$, $\vec{y}=\{y_{j}\}$ and the correlation matrix $T=t_{ij}$ can be obtained as:
\begin{align}
x_{i} &= \textmd{tr}\rho^{ab}(X_{i} \otimes I^{b})/\sqrt{n}\nonumber, \\
y_{j} &= \textmd{tr}\rho^{ab}(I^{a} \otimes Y_{j})/\sqrt{m}\nonumber, \\
t_{ij} &=\textmd{tr}\rho^{ab}(X_{i}\otimes Y_{j}).
\end{align}

Following the definitions made above, recently, an interesting analytical formula for the GMQD of an arbitrary two-qubit state has been introduced \cite{discord4}:
\begin{equation}
D_{G}(\rho^{ab})=2(\textmd{tr}S-\max\{k_{i}\}),
\end{equation}
where $S=\vec{x}\vec{x}^{t}+TT^{t}$ and
\begin{equation}
k_{i}=\frac{\textmd{tr}S}{3}+ \frac{\sqrt{6\textmd{tr}S^{2}-2(\textmd{tr}S)^{2}}}{3}\cos\left(\frac{\theta+\alpha_{i}}{3}\right),
\end{equation}
with $\{\alpha_{i}\}=\{0,2\pi,4\pi\}$ and $\theta=\arccos\{(2\textmd{tr}S^{3}-9\textmd{tr}S\textmd{tr}S^{2}+9\textmd{tr}S^{3})\sqrt{2/(3\textmd{tr}S^{2}
-(\textmd{tr}S)^{2})^{3}}\}$. Furthermore, observing that $\cos\left(\frac{\theta+\alpha_{i}}{3}\right)$ reaches its maximum for $\alpha_{i}=0$ and choosing $\theta$ to be zero, a very tight lower bound to the GMQD can be obtained, and it is given by:
\begin{equation}
Q(\rho^{ab})=\frac{2}{3}(2\textmd{tr}S-\sqrt{6\textmd{tr}S^{2}-2(\textmd{tr}S)^{2}}).
\end{equation}

We will refer to this quantity as the observable measure of quantum correlations (OMQC). It satisfies all of the criteria to be a meaningful measure of quantum correlations on its own. Moreover, it has the important advantage of requiring no optimization in the evaluation process. On top of being easier to calculate than the GMQD, it can be measured by performing seven local projections on up to four copies of the state. Thus, $Q(\rho)$ is also very experimentally friendly, since one does not need to perform a full tomography of the state.

\subsection{Measurement Induced Nonlocality}

After the introduction of the quantum discord, various quantum and total correlation measures have been proposed to quantify different kinds of correlations contained in a quantum system. Measurement-induced nonlocality (MIN) is a correlation measure that encapsulates the nonlocality contained in a bipartite system, in a more general way than the violation of Bell inequalities or entanglement, from a geometric perspective based on von Neumann measurements \cite{luo11}. The physical setting behind MIN is as follows: We perform von Neumann measurements on one of the subsystems of a bipartite quantum system and interpret the difference of the pre- and post-measurement states as a quantifier of nonlocality. The measurement operators are chosen so that they do not disturb the applied subsystem, in order to quantify solely the nonlocal effect of the measurement. Mathematically, it is defined by (taking into account the normalization):
\begin{equation} \label{min}
N(\rho^{ab})=2\max_{\Pi^{a}}\|\rho^{ab}-\Pi^{a}(\rho^{ab})\|^{2},
\end{equation}
where the maximum is taken over the von Neumann measurements $\Pi^{a}=\{\Pi_{k}^{a}\}$ that do not change $\rho^{a}$ locally, meaning $\sum_{k}\Pi_{k}^{a}\rho^{a}\Pi_{k}^{a}=\rho^{a}$, and $\|.\|^{2}$ denotes the square of the Hilbert--Schmidt norm. Although MIN, as given by Equation (\ref{min}), has no closed form formula for an arbitrary bipartite state, it can be calculated analytically for pure states of arbitrary dimension and for $2\times N$ dimensional mixed states. MIN for a $2\times 2$ dimensional system (two-spin system), which will be the focus of this work, can be analytically evaluated as:
\begin{align}
 N(\rho) = \begin{cases}
   2(\textmd{tr}TT^{t}- \frac{1}{\|\vec{x}\|^{2}} \vec{x}^{t}TT^{t}\vec{x}) & \text{if } \vec{x} \neq 0, \\
   2(\textmd{tr}TT^{t}- \lambda_{3}) & \text{if } \vec{x} = 0,
  \end{cases}
\end{align}
where $TT^{t}$ is a $3 \times 3$ dimensional matrix with $\lambda_{3}$ being its minimum eigenvalue, and $\|\vec{x}\|^{2}=\sum_{i}x_{i}^{2}$ with $\vec{x}=(x_{1},x_{2},x_{3})^{t}$. Due to the special form of the density matrix considered in this work, which is given in Equation (\ref{twospindm}), and since the local Bloch vector $\vec{x}$ is never zero in our investigation, MIN takes the simple form:
\begin{equation}
 N(\rho )=4(\rho_{23}^2+\rho_{14}^2).
\end{equation}

\subsection{Wigner--Yanase Skew Information}

Wigner--Yanase skew information (WYSI) mainly quantifies the information encapsulated in a quantum state with respect to an observable that is not commuting with it \cite{WYSI}. In recent years, WYSI has attracted a lot of attention in very different contexts, leading to different physical interpretations~ \cite{WYSIfisher1,WYSIfisher2,WYSIuncer1,WYSIuncer2}. The mathematical expression for WYSI is given as:
\begin{equation} \label{WYSI}
I(\rho, K)=-\frac{1}{2}\textmd{Tr}[\sqrt{\rho},K]^2,
\end{equation}
where $K$ denotes an observable, \textit{i.e}., a self-adjoint matrix. A very important property of WYSI is that it captures the genuine quantum uncertainty of a given observable in a certain quantum state. In other words, it does not get affected by the classical mixing. The basic properties of the WYSI can be listed as follows:
\begin{itemize}
\item It is upper bounded by the variance with respect to the considered observable, $I(\rho, K)\leq V(\rho,K)=\textmd{Tr}\rho K^2-(\textmd{Tr} \rho K)^2$, with the equality reached for pure states.

\item $I(\rho, K)$ is convex, such that:
\begin{equation}
I\left(\sum_i\lambda_i\rho_i, K \right)\leq \sum_i\lambda_iI(\rho_i, K),
\end{equation}

where $\rho_i$ is an arbitrary quantum state and $\lambda_i$ is a probability distribution satisfying the constraint $\sum_i\lambda_i=1$.

\item For a given bipartite state $\rho^{ab}$ and an observable $K$ acting on the subsystem $a$, one has:
\begin{equation}
I(\rho^{ab}, K\otimes I)\geq I(\rho^a, K).
\end{equation}

\end{itemize}

Coherence is one of the most significant properties that separates quantum states from classical ones. Although various different intuitive methods to measure the coherence of a given quantum state are known, there were no general frameworks for quantifying coherence. Recently, a list of defining properties for the coherence measures has been introduced in order to properly quantify coherence \cite{cohmon}. These properties can be stated as follows: a coherence monotone is: (i) zero if and only if the state is incoherent; (ii) non-increasing under incoherent operations; and (iii) non-increasing under classical mixing of quantum states. It has been shown that WYSI satisfies all of the criteria for coherence monotones \cite{WYSIcoh} and is defined as a $K$-coherence measure, meaning that the quantifier of coherence is contained in $\rho$ when measuring the observable $K$. In fact, the intuitive physical explanation behind this can be expressed as follows: WYSI gives information only about the quantum uncertainty appearing as a result of a measurement, and this uncertainty has its root directly in the coherence embodied in a state. Here, we follow this line of work and adopt WYSI as a coherence measure, referring to it as local quantum coherence (LQC).

Furthermore, it is possible to obtain a lower bound for WYSI by dropping the square root on the density matrix:
\begin{equation} \label{simpleWYSI}
I^L(\rho, K)=-\frac{1}{4}\textmd{Tr}[\rho,K]^2.
\end{equation}

This lower bound is particularly important due to its easy experimental accessibility. It can be measured using two programmable measurements on an ancillary qubit. Note that LQC can be calculated both for single and bipartite systems. Naturally, for bipartite systems, it is written as $I(\rho_{AB}, K_{A}\otimes I_{B})$, which quantifies the coherence with respect to a local observable acting on the first subsystem.

We have already mentioned that WYSI does not get affected by the classical mixedness of the quantum state. Motivated by this property of WYSI, in \cite{WYSIqc}, the local quantum uncertainty (LQU) of a given state with respect to an observable is defined as the minimum achievable WYSI on a single local measurement:
\begin{equation} \label{lqc}
U_A^\Gamma=\min_{K_A^\Gamma}I(\rho,K_A^\Gamma),
\end{equation}
where $\Gamma$ denotes the spectrum of $K_A^\Gamma$, and the minimization over a chosen spectrum of observables leads to a specific measure from the family. However, for a two-qubit system, all of the members of the family turn out to be equivalent. Then, the LQU can be analytically calculated as:
\begin{equation} \nonumber
U_A(\rho_{AB})=1-\lambda_{\max}\{W_{AB}\},
\end{equation}
where $\lambda_{\max}$ is the maximum eigenvalue of the $3\times 3$ symmetric matrix $W_{AB}$, whose elements are given~by:
\begin{equation} \nonumber
(W_{AB})_{ij}=\textmd{Tr}\left\{\sqrt{\rho_{AB}}(\sigma_{iA}\otimes I_B)\sqrt{\rho_{AB}}(\sigma_{jA}\otimes I_B)\right\},
\end{equation}
where indices $i,j=\{x,y,z\}$ are given for the usual Pauli operators. We note that Equation (\ref{lqc}) is normalized to one for maximally entangled pure states and, moreover, reduces to the linear entropy for any pure bipartite state.

Lastly, we introduce a measure for total correlations based on the WYSI, which we will refer to as the Wigner--Yanase skew information-based measure (WYSIM) \cite{WYSIcorr}. As $I(\rho,X)$ depends both on the state $\rho$ and the observable $X$, an averaged quantity has been introduced in order to get an intrinsic expression:
\begin{equation}
Q(\rho)=\sum_{i}I(\rho,X_{i}),
\end{equation}
where $\{X_{i}\}$ is a family of observables, which constitutes an orthonormal basis. It is possible to use $Q(\rho)$ to capture the total information content of a bipartite quantum system $\rho^{ab}$ with respect to the local observables of the subsystem $a$ as follows:
\begin{equation}
Q_{a}(\rho^{ab})=\sum_{i}I(\rho^{ab},X_{i}\otimes I^{b}),
\end{equation}
which is independent of the choice of a specific orthonormal basis $\{X_{i}\}$. As a result, the difference between the information content of the composite system $\rho^{ab}$ and the Kronecker product of the local subsystems $\rho^{a} \otimes \rho^{b}$ with respect to the local observables of the subsystem $a$ can be introduced as a measure for total correlations for $\rho^{ab}$,
\begin{align}
F(\rho^{ab}) &= \frac{2}{3}(Q_{a}(\rho^{ab})-Q_{a}(\rho^{a}\otimes\rho^{b})) \nonumber, \\
  &= \frac{2}{3}(Q_{a}(\rho^{ab})-Q_{a}(\rho^{a})),
\end{align}
where we add a normalization factor $2/3$.

Although many entanglement and other correlation measures mostly involve complicated optimization procedures, WYSI enjoys the advantage of being able to be calculated quite straightforwardly. Another entropic measure for total correlations, which is also easy to calculate, is the quantum mutual information introduced in Equation (\ref{mi}), which has also been shown to detect the CPs of various quantum critical systems. At this point, it is important to note that quantum mutual information and WYSI have different physical interpretations and reveal different information about the spin chain under consideration, which, in return, extends our understanding about the behavior of quantum information in such systems.

\subsection{Concurrence}

In order to quantify the entanglement content of a two-qubit density matrix, we utilize concurrence, which is the most common way to measure entanglement for these systems \cite{concur1,concur2}. First, we need to calculate the time-reversed or spin-flipped density matrix $\tilde{\rho}$, which is given by:
\begin{equation} \nonumber
\tilde{\rho}=(\sigma^{y}\otimes\sigma^{y})\rho^{*}(\sigma^{y}\otimes\sigma^{y}).
\end{equation}
Here, $\sigma^{y}$ is the Pauli spin operator and $\rho^{*}$ is obtained from $\rho$ via complex conjugation. Then, concurrence reads:
\begin{equation}
C(\rho)=\max \left\{ 0,\sqrt{\lambda_{1}}-\sqrt{\lambda_{2}}-\sqrt{\lambda_{3}}-\sqrt{\lambda_{4}},\right\},
\end{equation}
where $\{\lambda_{i}\}$ are the eigenvalues of the product matrix $\rho \tilde{\rho}$ in decreasing order. For the simple form of the reduced density matrix given in Equation (\ref{twospindm}), concurrence reduces to:
\begin{equation}
 C=2\max\{0, |\rho_{14}|-|\rho_{22}|, |\rho_{23}|-\sqrt{\rho_{11}\rho_{44}}\}.
\end{equation}

\section{Behavior of Correlations}

The investigation of various quantum correlation measures in the ground states of spin models have been an active area of research in the last decade. Before this line of research was initiated, Preskill argued that better understanding of entanglement in strongly-interacting many-body systems may deepen our knowledge of these systems \cite{preskill}. Following this idea, one of the first steps was taken in the direction of analyzing the single-site and two-site entanglement in the anisotropic XY chain in the transverse magnetic field (and in a limit of the same model, the transverse Ising chain) \cite{et01}. In this work, along with many important results, it has been shown that the quantum critical point in these systems can be identified by looking at the behavior of the entanglement in the ground state of the model, as quantified by concurrence and von Neumann entropy. Both measures are calculated for nearest-neighbor or next nearest-neighbor spins, and it has been concluded that the bipartite entanglement close to the CP in these systems is short-ranged. Moreover, the study includes the exploration of the bipartite entanglement in finite temperatures. It can also be seen that the entanglement is not very susceptible to the effect of temperature. In \cite{et02}, entanglement in the finite-sized XY chain is investigated. It has been shown that it obeys a scaling behavior near the critical point of the QPT in the model. In order to understand the connection between the non-analyticities of the bipartite entanglement measures and the QPTs, a general framework has been introduced. In \cite{wu}, it has been discussed that the behavior of bipartite entanglement measures is directly related to discontinuities occurring in the ground state energy. Therefore, the conditions to detect QPTs are obtained under the assumptions that these discontinuities are not accidentally canceled or no artificial discontinuity is created in the optimization procedure involved in the calculation of the measure. The last assumption is particularly important, because there are some explicit examples in the literature where such cases occur \cite{et011}.

On the other hand, the behavior of multipartite entanglement in spin chains has also been investigated. In \cite{et014,et015,et016}, based on a multipartite entanglement measure, which is a generalization of a global entanglement measure \cite{meyer}, it has been demonstrated that it is also possible to find the signatures of the QPT in some quantum critical models. Moreover, the multipartite entanglement has been determined to be larger than the amount of bipartite entanglement in the vicinity of QPT, suggesting that the entanglement is distributed over the whole lattice near criticality. The authors also have extended the framework introduced in \cite{wu} to the case of multipartite entanglement. Recently, in \cite{et017} and in \cite{et018}, multiparticle entanglement in anisotropic XY model in transverse field have been explored using different criteria for detecting the entanglement. Both papers show that in spite of the lack of bipartite entanglement, multipartite entanglement is present nearly for all values of the external field and the anisotropy parameter. Furthermore, they show that both entanglement criteria obey the finite-size scaling behavior of the XY model.

We have already mentioned that the symmetry breaking in the ground state of the considered spin chain models is generally not taken into account in the majority of the studies on this subject. However, there are several works exploring the effects of SSB on entanglement near criticality. The initial work in this direction was made in \cite{ssb1}, where it has been shown that the symmetry breaking does not effect entanglement in the ground state of the XXZ%please define
 and Ising model in general. Indeed, through an inequality involving the correlation functions of the system introduced in this work, it has been possible to determine whether the symmetry breaking will change the amount of entanglement. Later, in \cite{ssb2}, the work of \cite{ssb1} was extended to the case of the XY model in transverse field and the Lipkin--Meshkov--Glick model, and it has been demonstrated that for both of these models, there has been an enhancement in the entanglement when the symmetry breaking effect has been taken into account. Furthermore, multipartite entanglement has been shown to be affected more severely than the bipartite entanglement by the SSB near the quantum critical point \cite{ssb3,ssb4}.

Many important studies followed this line of work, investigating the interplay between entanglement and QPTs in different spin chain models, such as XXZ, XYZ%please define
, XY with the Dzyaloshinskii--Moriya (DM) interaction, Lipkin--Meshkov--Glick (LMG), \textit{etc}., both in the thermodynamic limit and in the few-body~cases.

Apart from the relation between the QPTs and entanglement, another important question is the dynamics of entanglement in the quantum spin chain, which was also addressed \cite{et06}. Here, the authors analyze how an initially entangled bipartite state evolves under the dynamics governed by the XY Hamiltonian as a function of the system parameters and the distance between the initially prepared~spins.

Now, we turn our attention to the relationship between QPTs and quantum correlation measures that are more general than entanglement. The interest in this subject has increased after the introduction of quantum discord, which is the first and most widely explored measure of quantum correlations beyond entanglement. Following quantum discord, many other measures have been introduced, some of which we will also explore in the context of the XY model.

The pioneering work on the investigation of general quantum correlations in spin chains was made in~\cite{dt01}. In this work, the author considered the transverse Ising chain together with the antiferromagnetic XXZ chain and analyzed the behavior of quantum discord for two nearest-neighbor spins, close to the CP of both systems. It has been shown that the quantum discord gets larger close to the CP for both of the considered models, but does not become maximum in the Ising model case. Due to this abrupt increase, the derivative of quantum discord is able to signal the presence of the QPT. Following this, a similar discussion to the one made in \cite{wu} has also been considered, and the relation between the discontinuities in the reduced density matrix elements and the quantum discord has been put forward. On the other hand, while the classical correlations decrease as one gets close to the QPT for the XXZ chain, they show a monotonically increasing behavior in the Ising model. A generalization of this work has been made in \cite{dt02}, which gives a general recipe for the calculation of quantum discord in states possessing $Z_2$ symmetry, a general spin flip symmetry exhibited by most of the spin chain Hamiltonians, but not necessarily exhibited by the ground state density matrix in the ordered phase, as we have mentioned. In order to illustrate the findings, the author has investigated the behavior of correlations in XXZ, Ising and LMG models, both in the thermodynamic limit and in the finite size cases. He has shown that the effects of first-, second- and infinite-order QPTs in these systems can be found in the classical correlations and quantum discord, either via the measure's or its derivatives behaviors near criticality. Furthermore, the scaling law obeyed by the derivative of the quantum discord was shown to differ compared to that of the entanglement. A comprehensive study of the two-site scaling of quantum discord in the XXZ, XY and Ising models in a transverse magnetic field has been done in \cite{dt013}. It is important to note that the geometric version of the quantum discord has also been explored in the anisotropic XY model \cite{cheng}. \linebreak It has been shown that the derivative of the geometric quantum discord is also singular at the CP, together showing a universal finite-sized scaling behavior.

The analysis of quantum correlations at finite temperatures has also attracted much attention due to the contrast of the obtained results with respect to the ones obtained for entanglement measures. One of the major contributions on this subject has been made in \cite{werlang1,werlang2,werlang3}, in which the quantum discord in the anisotropic XY, XXZ and Ising models in a transverse magnetic field has been explored at both zero and finite temperatures. They have shown that through the behavior of bipartite, nearest-neighbor thermal quantum discord (TQD), which is merely the quantum discord calculated at finite temperature, it is possible to detect the position of the QPT in the parameter space of the model Hamiltonian at finite temperatures. The particular importance of this work also stems from the fact that entanglement was not able to signal the CP at non-zero temperatures, as well as the thermodynamic quantities, such as entropy, specific heat, magnetization and magnetic susceptibility. Since the exact zero temperature is not attainable in real physical systems, such a result proves the advantage of quantum discord in experimental systems. Furthermore, the ability in estimating the correct values of CPs (the ones at zero temperature) in these systems was compared using the quantum discord, entanglement and other thermodynamic quantities. It has been revealed that although at temperatures very close to zero, all of these quantities pinpoint the CPs with acceptable success, as the temperature increases, quantum discord outperforms other measures. In another work \cite{dt05}, the TQD has also been investigated in two limits of the anisotropic XY model in a transverse field; the XX and Ising models. The authors have shown that the quantum correlations in the system can increase with temperature in high magnetic fields. They suggest that the explanation behind this effect is: when the magnetic field is on, the low-lying excited states are more correlated than the ground state, and with increasing temperature, the system favors those excited states.

The investigation of correlations for distant bipartite spin pairs is also an important subtopic in the investigation of correlations in spin chains. We have already mentioned that the bipartite entanglement near criticality in the spin models is generally short ranged; beyond next nearest neighbors, it is not possible to find a highly entangled pair. However, this is not the case for quantum correlations that are more general than entanglement. In \cite{dt05,dt09}, it has been shown that in the XY model, in contrast to the short-ranged entanglement, as the system approaches the ordered phase, a long-ranged quantum discord develops in the system. A more systematic approach to the long-range correlations in the same model has been exhibited in \cite{nott05}, where both zero and finite temperature cases have been explored. For the zero temperature case, it has been demonstrated that the quantum discord of two-spins, which are 15 spins apart, was able to capture the QPT. Indeed, it behaves like an order parameter of the transition, being zero in the disordered phase, while becoming non-zero with a finite jump in the ordered phase. On the other hand, at finite temperatures, quantum discord still seems to be able to spotlight the QPT up to 15 spin separations; however, it becomes very fragile to the effects of temperature.

Similar to the case of multipartite entanglement, there has also been an interest in the definition, characterization and operational meaning of the QD generalized to multipartite states, which goes by the name, global quantum discord (GQD). A widely-accepted definition of GQD has been made in \cite{dt07} by a systematic extension of the original definition of bipartite QD, which is done by rearranging it in terms of relative entropy and local von Neumann measurements. The authors have also demonstrated the behavior and content of GQD in Werner-GHZ%please define
 states and the Ashkin--Teller spin chain. The latter possesses an infinite-order QPT in its phase diagram. It has been shown that GQD is able to capture the presence of this CP where bipartite QD is not able to do so. The results on GQD are extended to various spin chain models at both zero and finite temperatures \cite{nott02,nott03}. The behavior of GQD has been studied in finite-sized transverse Ising, cluster-Ising and XX models in a transverse magnetic field with open boundary conditions at finite temperature, and it was shown that the non-trivial changes in these systems can be captured by looking at the GQD. Furthermore, it has also been demonstrated that, for the Ising model, finite-sized scaling of the GQD is characterized by the universal critical exponents belonging to the Ising universality class.

The factorization phenomena that we introduce in Section 2 is not always so straightforward to detect with the help of the correlation measures. First of all, the point where the ground state of the system factorizes is always in the ferromagnetic (ordered) phase of the system, as can be seen from Equation~(\ref{factor}). Thus, in principle, in order to see a clear signature of the factorization on the correlation measures, one needs to consider the effects of symmetry breaking in the ground state of the system. However, even when SSB is ignored, bipartite entanglement measures (concurrence) for a fixed distance between two spins are able to detect it by being exactly equal to zero at this point. On the other hand, QD for the thermal ground state of the XY model can show the existence of this factorized point by the single crossing of the QD curves for different bipartite spin distances, \textit{i.e}., by having the same value independent of the spin-spin distance \cite{ssbandnott01}. In a recent work \cite{nott05}, a different approach has been taken in order to understand the nature of the factorization phenomena. The authors considered a finite-sized spin chain, looked at the energy spectrum and saw that right at the factorization field, $\lambda_f$, the energy levels of the ground and the first excited states cross, which are of opposite parity. What happens at this point is actually a transition between the different parity ground states. They have also examined the factorization phenomena at finite temperatures and seen that as the temperature increases, the factorization point widens and becomes a region of separability.

We have given an overview of the main results on the behavior of QD in different spin chains. However, there are many other important works done on this topic, such as examining the nonlocality and other measures of non-classicality in various spin models with more exotic interactions in both few-body systems and in the thermodynamic limit. An emerging subject in this field is to examine the correlations in systems possessing a topological phase transition. Topological phase transitions cannot be characterized by an order parameter or a broken symmetry in the system, therefore laying outside the scope of the traditional way of understanding phase transitions \cite{topology}. Thus, they are characterized by the change in some topological properties in the ground state of the system. Attacking this problem with the tools of quantum information theory has also given some important insights about the nature of the topological phase transitions \cite{topo1,topo2,topo3,topo4,topo5,topo6}.

In the following sections, we will extensively review our contributions to the investigation of correlation measures in spin chains, which involves the study of the correlation, coherence and uncertainty measures introduced in Section 3 in the one-dimensional anisotropic XY model in a transverse magnetic field, both at zero and finite temperatures, with a special emphasis on the detection of the factorization phenomena and ambiguities caused by the optimization procedure of the correlation~measures.

\subsection{Correlation Measures}

We begin with the discussion of the total correlation measures as quantified by WYSIM and MIN, which are, together with their derivatives, presented in Figure 1. The measures are calculated for nearest neighbors and plotted as a function of the external field $\lambda$ for $kT=$ 0, 0.1, 0.5 and $\gamma=$ 0.001, 0.5, 1. It can be seen that both measures behave in a very similar fashion for $\gamma=1$ and $\gamma=0.5$ at all temperatures. Their behavior differs in the case of $\gamma=0.001$, where WYSIM makes a sharper jump to a finite value than MIN right at the CP. We can also see that the correlation content of the two-spin reduced density matrix increases as a function of $\lambda$, meaning that the system is more correlated in the ferromagnetic phase. Right at the CP $\lambda_c=1$, both measures make a finite jump, resulting in a divergence in their first derivatives, which points to the existence of the second-order QPT of the considered system.

\begin{figure}[H]
\centering
\begin{tabular}{@{}cc@{}}
\includegraphics[scale=0.6]{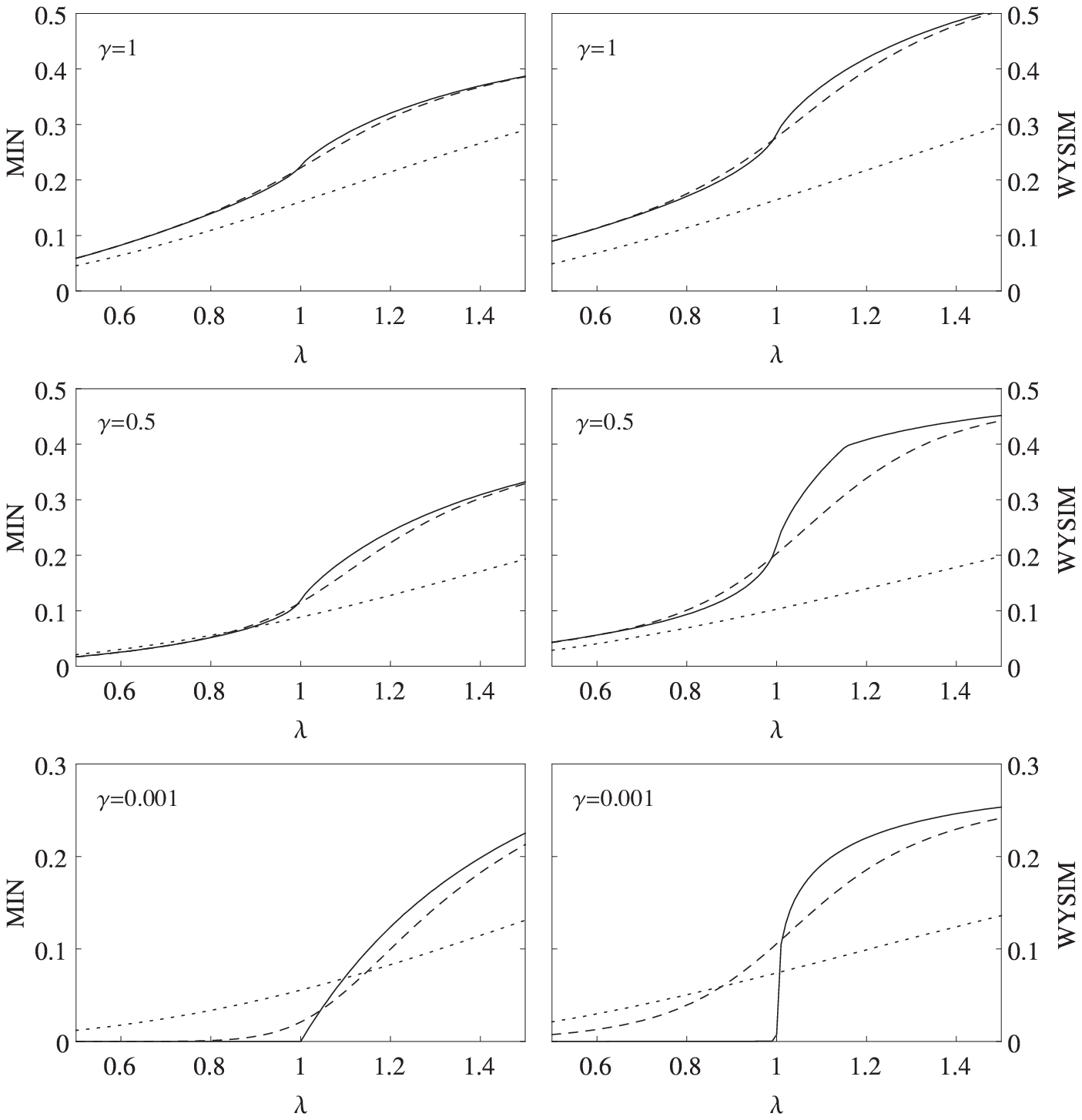} &
\includegraphics[scale=0.6]{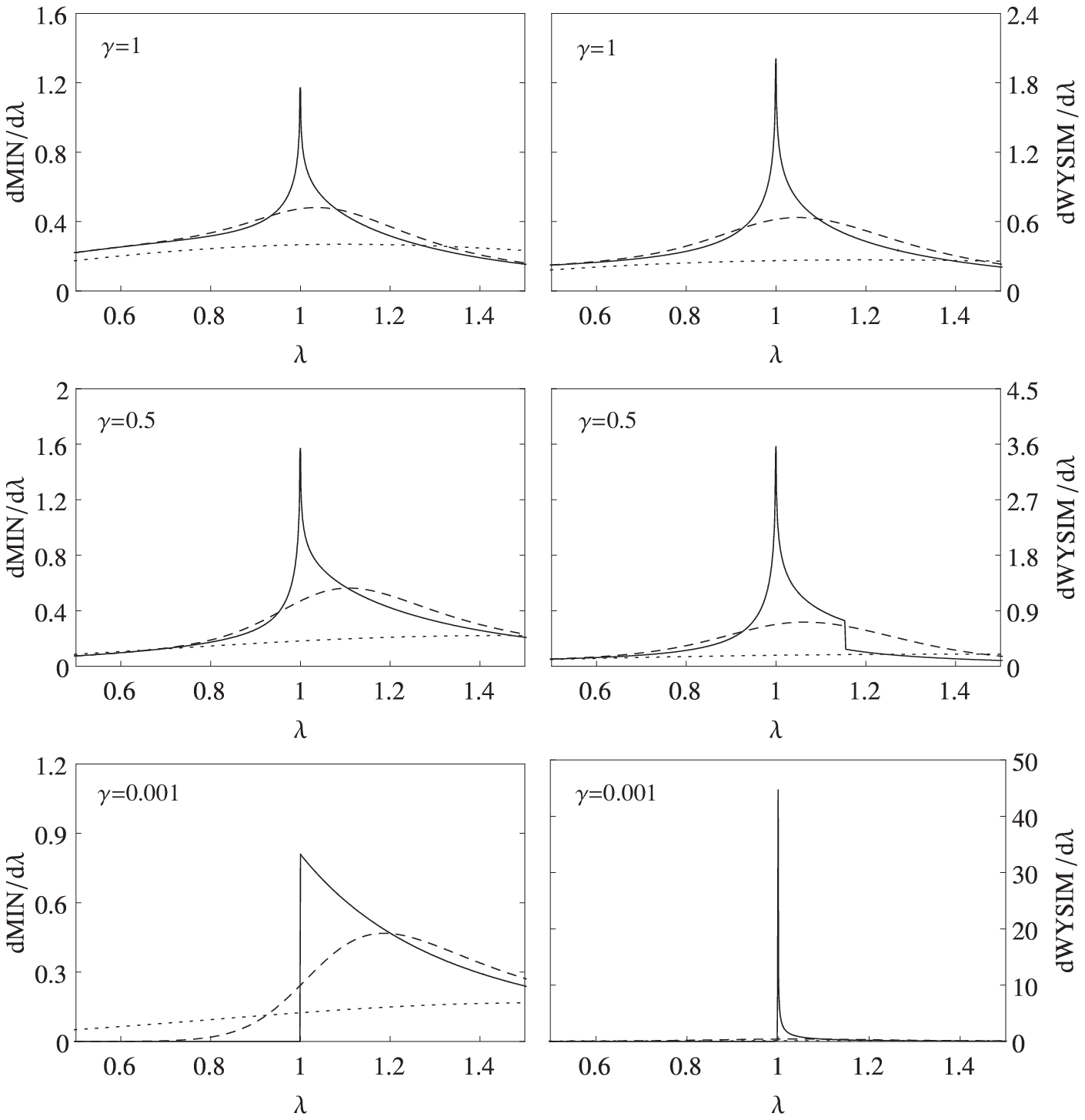} \\
(\textbf{a}) & (\textbf{b})
\end{tabular}

\caption{{\em Left panel} The thermal total correlations quantified by measurement-induced nonlocality (MIN) and the Wigner--Yanase skew information-based measure (WYSIM) as a function of $\lambda$ for $\gamma=$ 0.001, 0.5, 1 at $kT=0$ (solid line), $kT=0.1$ (dashed line) and $kT=0.5$ (dotted line). {\em Right panel} The first derivatives of MIN and WYSIM as a function of $\lambda$ for $\gamma=$ 0.001, 0.5, 1 at $kT=0$ (solid line), $kT=0.1$ (dashed line) and $kT=0.5$ (dotted line). The graphs are for the first nearest neighbors.}
\end{figure}
%\begin{figure}[H]
%\begin{center}
%\includegraphics[scale=0.56]{Totalfirst.eps}
%\caption{The thermal total correlations quantified by MIN and WYSIM as a function of $\lambda$ for $\gamma=0.001,0.5,1$ at $kT=0$ (solid line), $kT=0.1$ (dashed line) and $kT=0.5$ (dotted line). The graphs are for first nearest neighbors.}
%\end{center}
%\end{figure}
%
%\begin{figure}[H]
%\begin{center}
%\includegraphics[scale=0.56]{Totalfirstder.eps}
%\caption{The first derivatives of MIN and WYSIM as a function of $\lambda$ for $\gamma=0.001,0.5,1$ at $kT=0$ (solid line), $kT=0.1$ (dashed line) and $kT=0.5$ (dotted line). The graphs are for first nearest neighbors.}
%\end{center}
%\end{figure}

We can now turn our attention to the investigation of quantum correlations as quantified by OMQC and concurrence, and their derivatives, presented in Figure 2. As in the total correlation case, the graphs are for nearest neighbors and plotted as a function of the external field $\lambda$ for $kT=$ 0, 0.1, 0.5 and \linebreak $\gamma=$ 0.001, 0.5, 1. Again, we can see that both measures are capable of detecting the CP in the system though the divergence in the first derivative. It can be seen that concurrence is not very susceptible to the effects of temperature, since it remains zero nearly for all parameter values at high temperatures. We also note for $\gamma=0.001$ that the behavior of OMQC and concurrence is very similar to that of MIN and WYSIM, respectively. This suggests that at this parameter range, the correlations contained in the system are mainly quantum.
\begin{figure}[H]
\centering

\begin{tabular}{@{}cc@{}}
\includegraphics[scale=0.6]{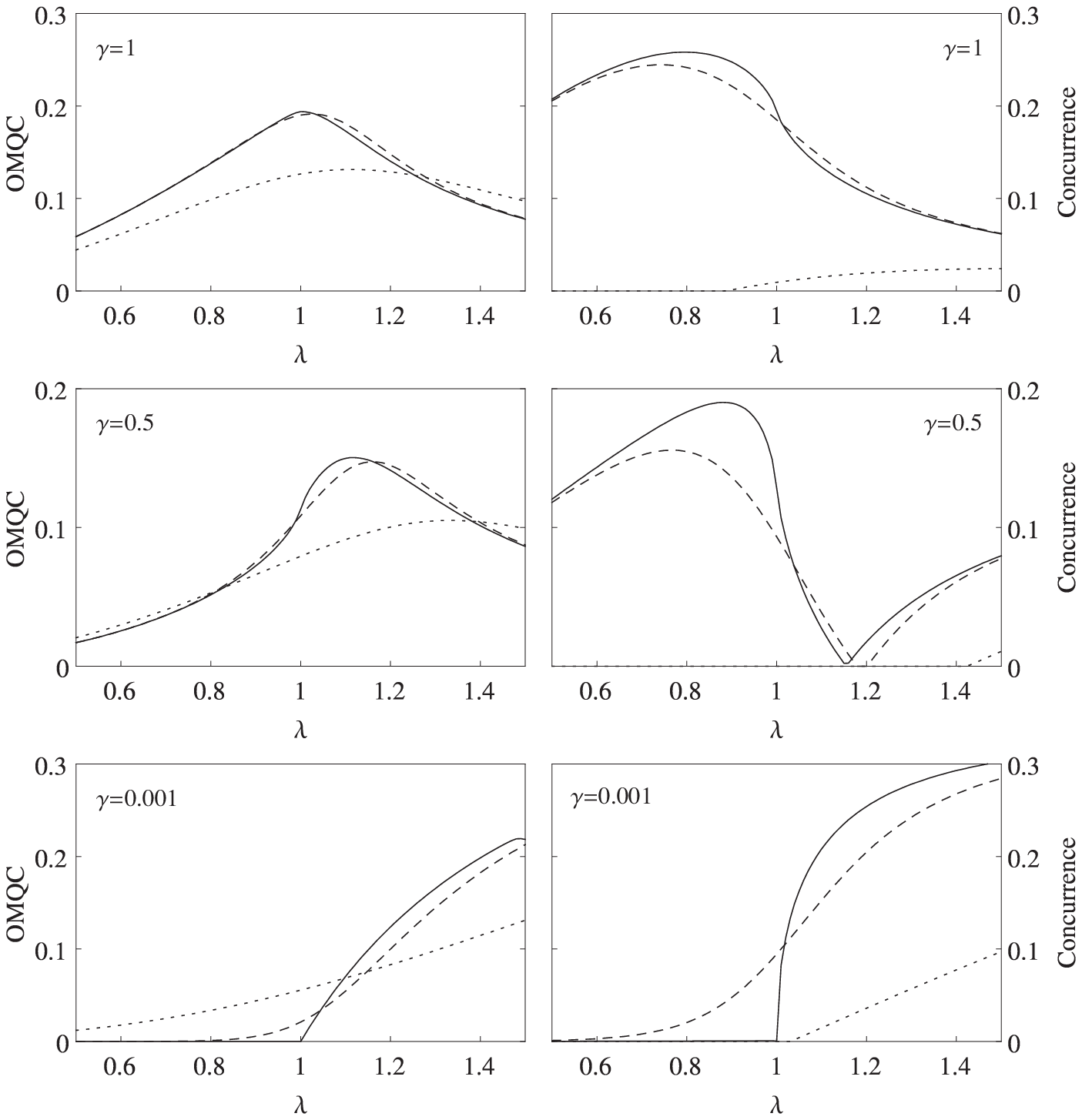} &
\includegraphics[scale=0.6]{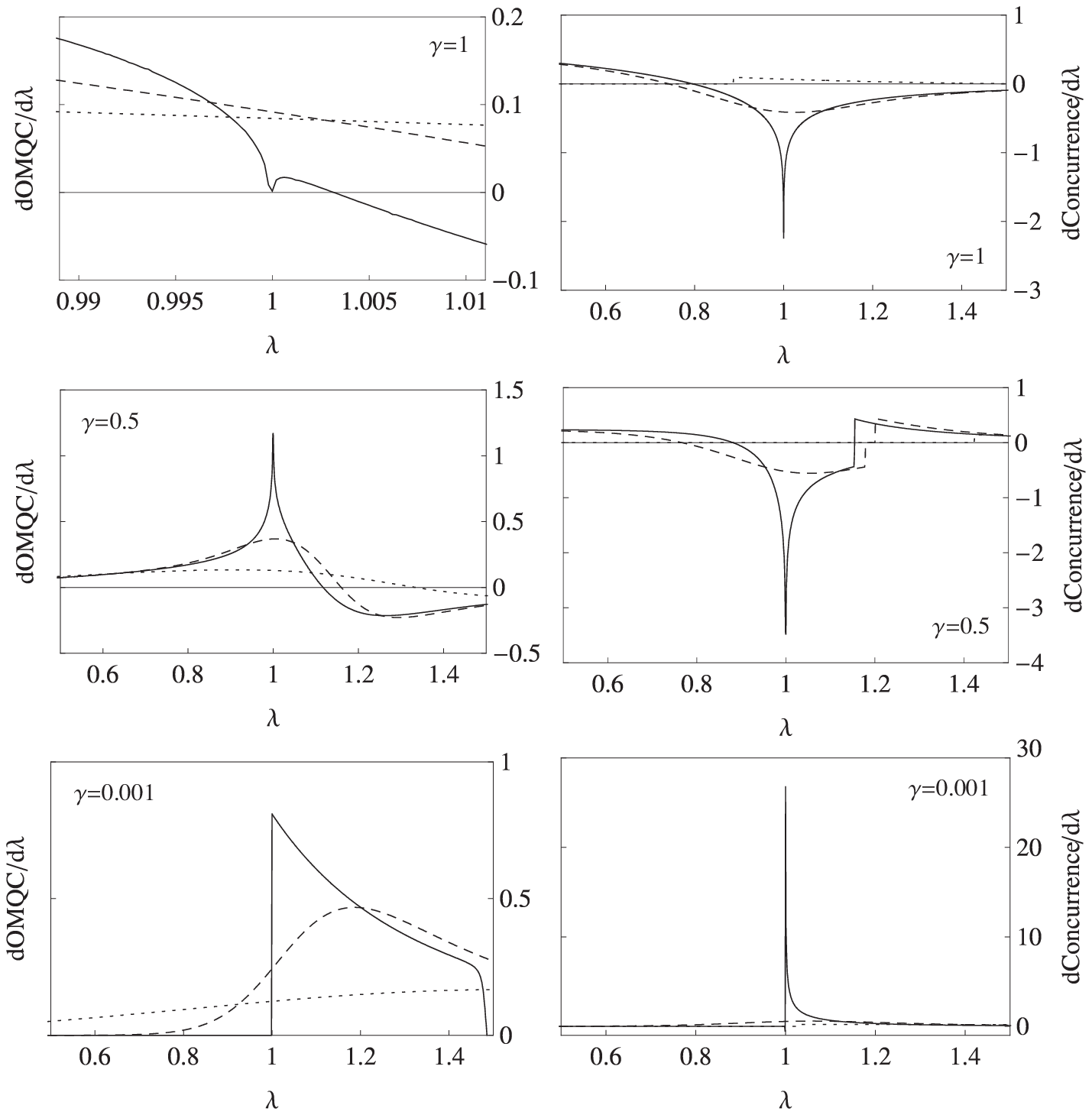} \\
(\textbf{a}) & (\textbf{b})
\end{tabular}

\caption{{\em Left panel} The thermal quantum correlations quantified by observable measure of quantum correlations (OMQC) and concurrence as a function of $\lambda$ for $\gamma=$ 0.001, 0.5, 1 at $kT=0$ (solid line), $kT=0.1$ (dashed line) and $kT=0.5$ (dotted line). {\em Right panel} The first derivatives of OMQC and concurrence as a function of $\lambda$ for $\gamma$ =0.001, 0.5, 1 at $kT=0$ (solid line), $kT=0.1$ (dashed line) and $kT=0.5$ (dotted line). The graphs are for first nearest neighbors.}
\end{figure}

Although the divergence of the derivatives of the measures is clearly pronounced for all values of $\gamma$ at $kT=0$, as the temperature increases, the finite jump of the measures near QPT smooths out, causing the divergences in the derivative to disappear. This is an expected result, since, as we have mentioned earlier, as temperature increases, thermal effects dominate the genuine quantum fluctuations. Additionally, similar results can be obtained if we look at the correlations for next nearest neighbors; however, they provide no additional information about the system, so we do not present them here.

Focusing on the signatures of the factorization phenomena, we first observe that for $\gamma=0.5$, concurrence becomes exactly zero at the factorization field $\lambda_f\approx 1.15$, which can be calculated from Equation (\ref{factor}). More interestingly, in all of the remaining three correlation measures considered in this work, only WYSIM is able to detect the existence of the factorized ground state by a non-analytic behavior in its derivative. This is a rather peculiar property of WYSIM, since, as mentioned, we do not take into account the effects of SSB. No other correlation measure in the literature is capable of revealing this phenomena without explicitly considering the effects of SSB with a single evaluation of the measure. The intersection point of the QD calculated using the thermal (symmetry protected) state for different spin distances can detect factorization in this model; however, for WYSIM, a single calculation is sufficient.

%\begin{figure}[H]
%\begin{center}
%\includegraphics[scale=0.56]{Quantumfirst.eps}
%\caption{The thermal quantum correlations quantified by OMQC and concurrence as a function of $\lambda$ for $\gamma=0.001,0.5,1$ at $kT=0$ (solid line), $kT=0.1$ (dashed line) and $kT=0.5$ (dotted line). The graphs are for first nearest neighbors.}
%\end{center}
%\end{figure}
%
%\begin{figure}[H]
%\begin{center}
%\includegraphics[scale=0.56]{Quantumfirstder.eps}
%\caption{The first derivatives of OMQC and concurrence as a function of $\lambda$ for $\gamma=0.001,0.5,1$ at $kT=0$ (solid line), $kT=0.1$ (dashed line) and $kT=0.5$ (dotted line). The graphs are for first nearest neighbors.}
%\end{center}
%\end{figure}

In Figure 3, we present our results on the performance of the measures in correctly estimating the CP at finite temperatures. We know that the divergent first derivative disappears and becomes a smoother function as the temperature increases. However, at sufficiently low temperatures, it is still possible to extract some information from the behaviors of the correlation measures in the vicinity of the QPT. The strategy we follow to estimate the CP at a finite temperature is to look for a local maximum or minimum at the first derivative and identify this point as the estimated CP. This extremum point can be identified easily by checking the second derivative of the measure, as is done here. A first glance, Figure 3 suggests that the accurateness of the estimation is strongly dependent on the parameters of the Hamiltonian.
\begin{figure}[H]
\begin{center}
\includegraphics[scale=0.75]{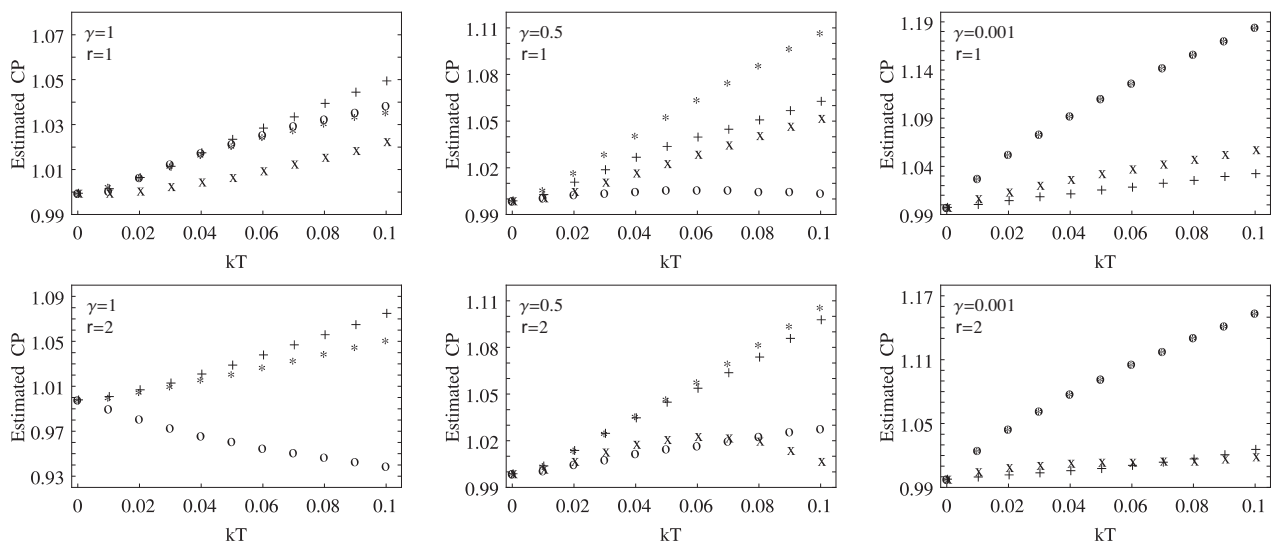}
\caption{The estimated values of the CP%please define
 as a function of $kT$ for three different values of the anisotropy parameter $\gamma=$ 0.001, 0.5, 1. The CPs in the graphs are estimated by OMQC (denoted by o), WYSIM (denoted by +), MIN (denoted by $\ast$) and concurrence (denoted~ by~x). Concurrence is not included for $\gamma=1$ and $r=2$, since it vanishes at even very low temperatures.}
\end{center}
\end{figure}

The graphs in the upper panel of Figure 3 are plotted for the nearest-neighbor spins for $\gamma=$ 0.001, 0.5, 1 as a function of temperature. The considered temperature range is from $kT=0$ to $kT=0.1$. In the case of $\gamma=1$, we see that the best and the worst estimators turn out to be the concurrence and WYSIM, respectively, with low relative error between them. OMQC and MIN nearly estimate the same value for CP, with MIN outperforming OMQC only very slightly as we approach $kT=0.1$. For~ $\gamma=0.5$, as OMQC shows a remarkable accuracy by approximately predicting the exact value of the CP at $kT=0$, MIN deviates from the true value rapidly and by a large amount. Lastly, when we change the anisotropy parameter to be $\gamma=0.001$, we observe that OMQC and MIN behave in the exact same manner and perform very poorly in the estimation process, while both concurrence and WYSIM predict the position of the CP better, with WYSIM being the closest to the original value. On the other hand, in the lower panel of Figure 3, the next nearest-neighbor spins for the same values of $\gamma$ and the same range of temperature as the upper panel are plotted. Again, starting with the the $\gamma=1$ case, we notice that WYSIM, MIN and OMQC lose their accuracy for the same amount, and the concurrence is not plotted for this case, since it quickly decays to zero, even for very small values of $kT$. Note that in all cases considered in Figure 3, only for these values of parameters do we see a measure, OMQC, estimating the CP as less than its true value. Considering the $\gamma=0.5$, the total correlation measures (WYSIM and MIN) perform worse than the quantum correlation measures (OMQC concurrence). Finally, for $\gamma=0.001$, we observe a very similar pattern with the graph plotted for the nearest neighbors.

A similar analysis that we presented here concerning the CP estimation with correlation measures at finite temperature has also been considered in another study, where the entanglement of formation (EOF) and QD are the subjects of the investigation. As compared to the performance of EOF and QD on the same task, it is not possible to make an absolute comment on the success or failure of the measures considered in this work, due to its high dependence on the system parameters. However, relative comparisons can be made, for example in the case of nearest neighbors with $\gamma=0.5$ OMQC or in the case of next nearest neighbors with $\gamma=0.001$ WYSIM, outperforming both EOF and QD.

We continue our discussion with the dependence of correlations on the distance between the spin pair for which they are calculated. It is known that bipartite entanglement is very short ranged in contrast to the QD. In Figure 4, we demonstrate our findings on the long-range behavior of WYSIM, MIN and OMQC for $\lambda=$ 0.75, 0.95, 1.05, 1.5, $\gamma=$ 0.001, 1 and $kT=$ 0.1, 0.5. It can be seen from the plots that temperature have a diminishing effect on the correlations between distant spins. We also show that for $\gamma=0.001$, all measures decay to zero as the distance between spin pairs are increased. On the other hand, in the Ising limit of our model, $\gamma=1$, correlations between distant spin pairs settle at a finite value for $\lambda=1.5$, which is deep in the ferromagnetic (ordered) phase. In fact, in all cases considered here, long-range correlations persist longer in the ordered phase.

\begin{figure}[H]
\begin{center}
\includegraphics[scale=0.75]{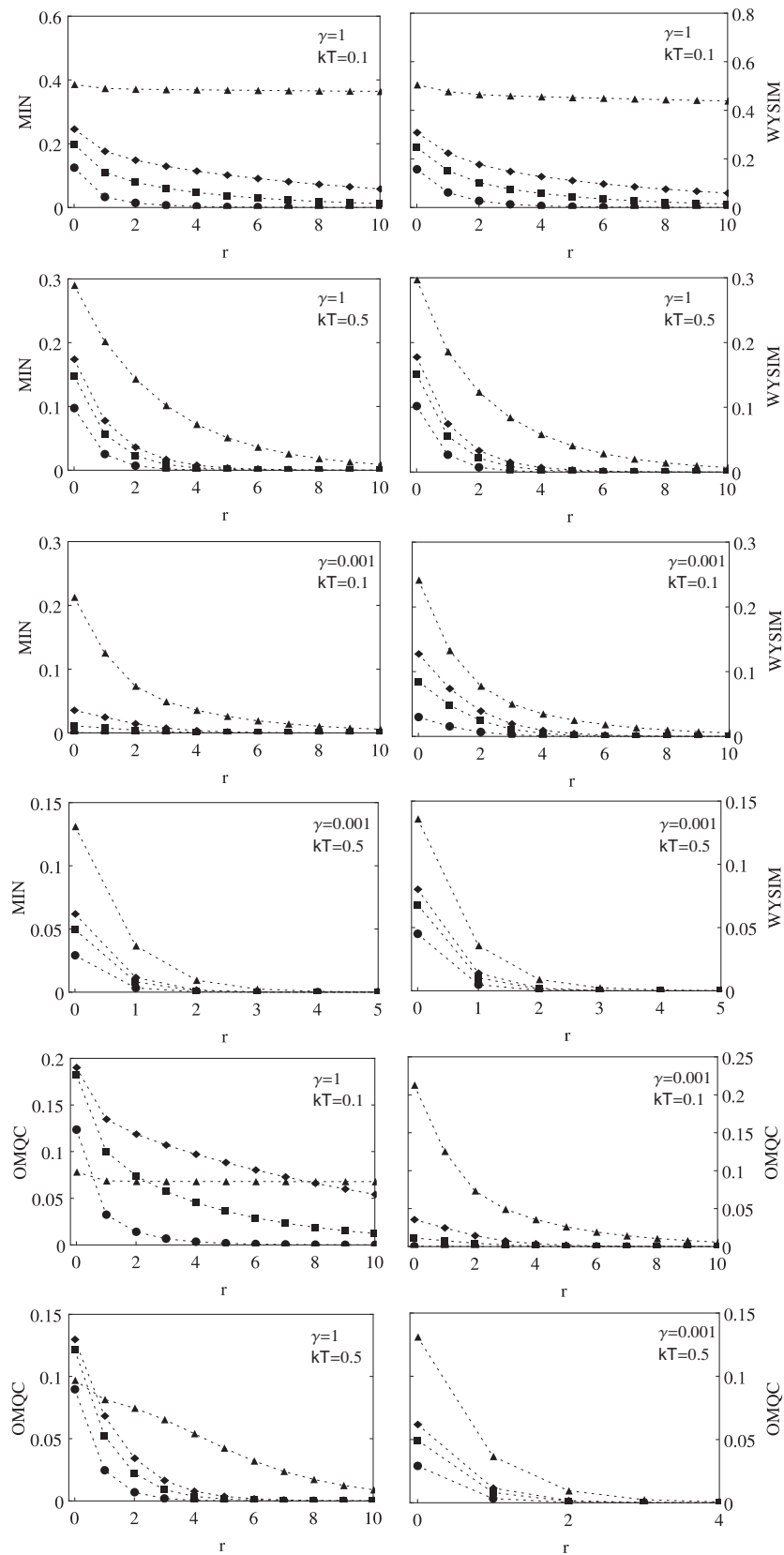}
\caption{Long-range behavior of the thermal total and quantum correlations for $\gamma=0.001$ and $\gamma=1$ at $kT=$ 0.1, 0.5. The circles, squares, diamonds and triangles correspond to $\lambda=0.75$, $\lambda=0.95$, $\lambda=1.05$ and $\lambda=1.5$, respectively. }
\end{center}
\end{figure}

\subsection{Local Quantum Coherence}

As introduced in the correlations section, LQC quantifies the coherence contained in a given density matrix with respect to a fixed basis. The coherence measure that we adopt in this work is merely the WYSI. We start by the discussion of single-spin $\sigma_x$ coherence (coherence contained in $\rho$ when measuring~$\sigma_x$), $I(\rho_0$, $\sigma_x)$ and its experimentally-friendly lower bound, $I^L(\rho_0$, $\sigma_x)$ in our system, which are presented in Figure 5. We plot these quantities for $\gamma=0.5$ and $\gamma=1$, which is the Ising model in transverse field.

At this point, it is important to note that we have also calculated the coherence with respect to different observables for both single- and two-spin cases; however, since they give no additional information about the system under consideration, we only show $\sigma_x$ coherence as a representative.

The left panel in Figure 5 shows the measures themselves, and the right panel displays their derivative. First, we recognize the fact that the single-spin $\sigma_x$ coherence and its lower bound decreases as the inverse field $\lambda$ increases, meaning that in the ordered phase, a randomly chosen spin contains less coherence than it contains the disordered phase. It can be seen from Figure 5b,d that the derivatives of the measures are capable of detecting the presence and the order of the CP by displaying a divergent behavior at $\lambda_c=1$. On the other hand, we see no non-trivial behavior at the factorization field $\lambda_f\approx 1.1547$ for $\gamma=0.5$, \textit{i.}e., the coherence of a single spin behaves smoothly at this point. We do not expect to see any effect of factorization for $\gamma=1$, since in this case, the factorizing field tends to infinity, $\lambda\rightarrow\infty$. It is remarkable and important that even the experimentally-friendly, simplified version of LQC, $I^L(\rho_0$, $\sigma_x)$, which does not require the full tomography of the state, can spotlight the CP of the QPT.

\begin{figure}[H]
\begin{center}
\includegraphics[width=0.7\textwidth]{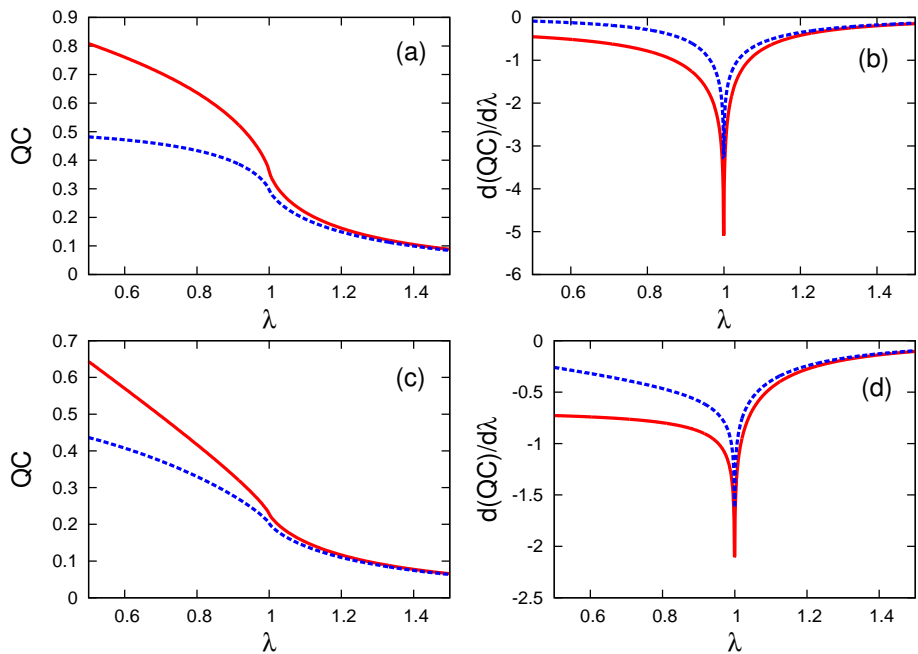}
\caption{Single-spin $\sigma_x$-coherence for $\gamma=0.5$ (\textbf{a}) and $\gamma=1$ (\textbf{c}), along with its first derivative (with respect to $\lambda$) for $\gamma=0.5$ (\textbf{b}) and $\gamma=1$ (\textbf{d}), as a function of $\lambda$. As the red solid line denotes the measure, the dashed blue line corresponds to its simplified version.}
\label{fig1}
\end{center}
\end{figure}

Our results on two-spin $\sigma_x$ LQC are presented in Figure 6, again for two different values of $\gamma$, $\gamma=~0.5$ and $\gamma=1$. Recall that in the two-spin case, we calculate the coherence contained in one of the subsystems locally. Therefore, the measurement operator only acts on the chosen subsystem, and the mathematical expression for this is given as $I(\rho_{AB}, \sigma_x\otimes I_{B})$. We have considered nearest neighbor spins in our discussion; however, similar results can also be obtained considering next-nearest neighbor spins. In the left panel, we can see the behavior of the measures and conclude that they seem very similar to that of the single-spin coherence. Both the original definition and its lower bound follow a decreasing trend with increasing inverse field. The derivatives of the measures, presented in the right panel, again signal the existence of the CP via a divergence in their derivatives, thus also giving the correct information about the order of the phase transition. The important difference from the single-spin case is the appearance of a small discontinuity in the original measure for $\gamma=0.5$ at $\lambda=1.1547$. As we have mentioned, this is the critical field in order to observe the factorized ground state of the system for the considered value of $\gamma$. While the WYSI can detect the occurrence of this phenomenon, the lower bound for WYSI does not get affected by it. It is rather striking that WYSI itself can signal the factorized ground state, since neither is it an entanglement measure for mixed states, nor have we considered the effects of symmetry breaking on the ground state of the system.
\begin{figure}[H]
\begin{center}
\includegraphics[width=0.7\textwidth]{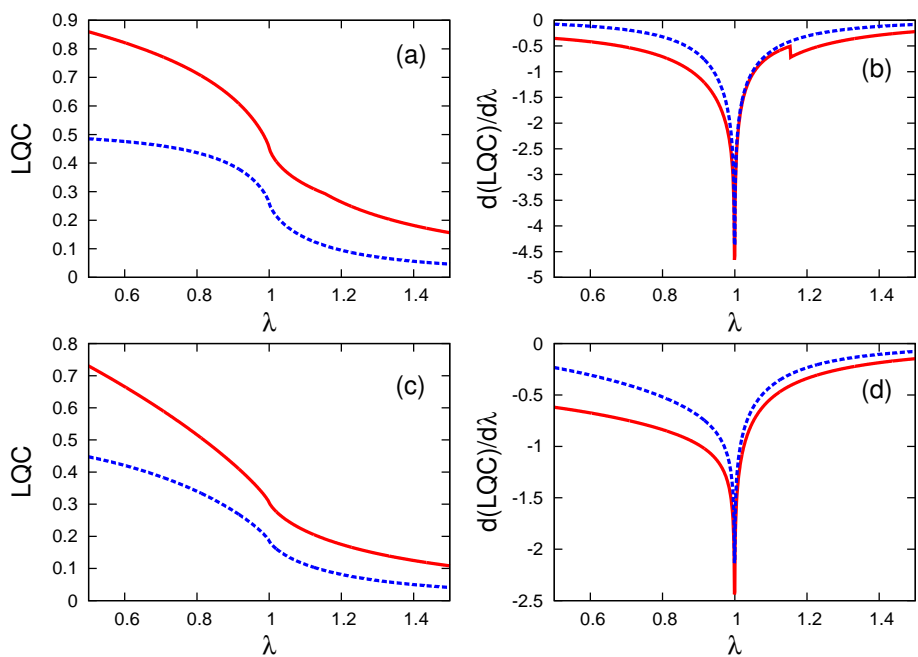}
\caption{Two-spin local $\sigma_z$-coherence for $\gamma=0.5$ (\textbf{a}) and $\gamma=1$ (\textbf{c}), along with its first derivative (with respect to $\lambda$) for $\gamma=0.5$ (\textbf{b}) and $\gamma=1$ (\textbf{d}), as a function of $\lambda$. As the red solid line denotes the measure, the dashed blue line corresponds to its simplified version.}
\label{fig3}
\end{center}
\end{figure}

Now, we will have a deeper discussion of the occurrence of discontinuities in the derivatives of WYSI and WYSIM at $\lambda_f$. First of all, it is important to note that these discontinuities are present for all values of $\gamma$, not just for the $\gamma=0.5$ case. The explanation of such a behavior stems from the presence of the square root of the density matrix, $\sqrt{\rho^{0r}}$, in the definition of WYSI. The elements of $\sqrt{\rho^{0r}}$ are themselves discontinuous at the factorization point, resulting in a physical quantity depending on them to be also discontinuous at the same point. Only correlation measures other than WYSI that can identify the factorization phenomenon in a single evaluation are entanglement measures, such as concurrence and the entanglement of formation, which is itself a function of concurrence. Interestingly, looking at the definition of concurrence, we see that it also involves $\sqrt{\rho^{0r}}$ and vanishes at $\lambda_f$ as a result of the fact that even the thermal ground state is factorized at this point. Furthermore, as soon as we drop the square root of the density matrix, for example as in the case of $I^L$, we suddenly lose the information on the ground state factorization. We stress that the detection mechanism for the factorization point is not the same as the detection of the CP, since no thermodynamic quantity is discontinuous at this point. To be clear, we do not present an explicitly constructed correspondence between the discontinuities of the elements of $\sqrt{\rho^{0r}}$ and the emergence of the factorization point; instead, we point to a possible explanation of the results that we have obtained.

In this last part of this section, we carry out the same investigation that we have done for the correlation measures in the previous section and look at the performance of LQC and its lower bound in correctly estimating the CP of the QPT and the factorization point. We have already explained that at temperatures low enough to see the effects of quantum fluctuations, it is possible to see the signatures of the CP, this time not through the divergences in the derivatives, but rather the extremum points near the CP. Similarly, at finite temperatures, the factorization point ceases to be a point and becomes a region of separability in the parameter space of the model. Finite temperature analysis is particularly important also for the reason that zero temperature is not attainable experimentally.

In Figure 7a, we plot the estimated CP as pointed out by the $\sigma_x$ LQC and its simplified version for $\gamma=0.5$ as a function of temperature. We see that the experimentally-friendly lower bound remains quite accurate up to a temperature $kT=0.1$. This is particularly important, since it proves that $I^L$ can be a strong candidate to detect the CP in experimental applications. Figure 7b shows the estimated factorization point calculated from the non-trivial behavior of $\sigma_x$, $\sigma_y$ and $\sigma_z$ coherence for $\gamma=0.5$ as a function of temperature. All three measures behave very similarly as the temperature increases. They stay extremely accurate until $kT=0.015$ and start to deviate significantly from the actual point from $kT=0.02$

\begin{figure}[H]
\begin{center}
\includegraphics[width=0.7\textwidth]{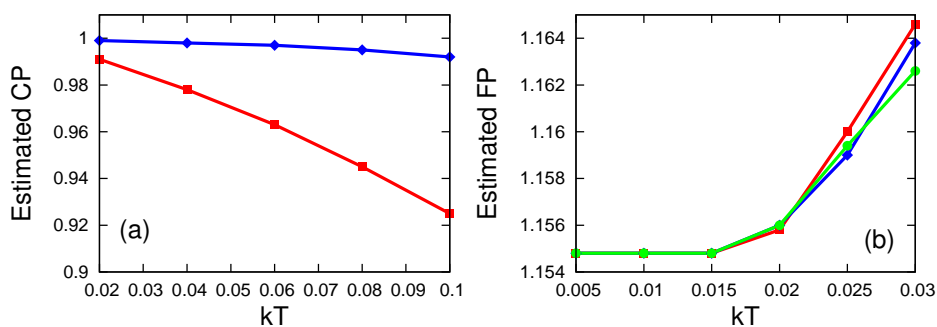}
\caption{(\textbf{a}) The critical point estimated by single-spin $\sigma_x$-coherence (red line) and its simplified version (blue line) as a function of the temperature for $\gamma=0.5$. (\textbf{b}) The factorization field estimated by local two-spin $\sigma_x$-coherence (red line), $\sigma_y$-coherence (blue line) and $\sigma_z$-coherence (green line) as a function of time for $\gamma=0.5$.}
\label{fig6}
\end{center}
\end{figure}
\newpage
\subsection{Local Quantum Uncertainty}

Lastly, we would like to discuss the results that we obtain on LQU, which is nothing but the optimized version of the LQC over all possible observables. In Figure 8, we plot LQU and its derivative for $\gamma=0.5$ and $\gamma=1$ as a function of $\lambda$. The measure itself has a small asymmetry around the peak point near $\lambda=1$, settling to a higher value in the ordered phase. Through the divergence at $\lambda_c=1$ and finite discontinuity at $\lambda_f=1.1547$, in its derivative, we can spot the presence of the CP and the factorization point, respectively. However, in Figure 8a,c, there are two extremum points, resulting in discontinuities in the derivatives, at the Hamiltonian parameter values where the system exhibits no non-trivial behavior. The reason behind these non-analyticities in the derivative of LQU is, in fact, due to the optimization procedure involved in the calculation of LQU. An abrupt change in the optimizing observable causes such extremum points in the measure itself. Specifically, in this case, two extremum points of LQU exactly correspond to the values of $\lambda$, where the optimizing observable changes from $\sigma_z$ to $\sigma_x$. Correlation measures involving optimization procedures might sometimes cause such ambiguities and display finite discontinuities that are not rooted in the elements of the reduced density matrix. Therefore, when dealing with such measures, one should be careful not to confuse a non-analyticity resulting from the optimization with a QPT.

\begin{figure}[H]
\begin{center}
\includegraphics[width=0.7\textwidth]{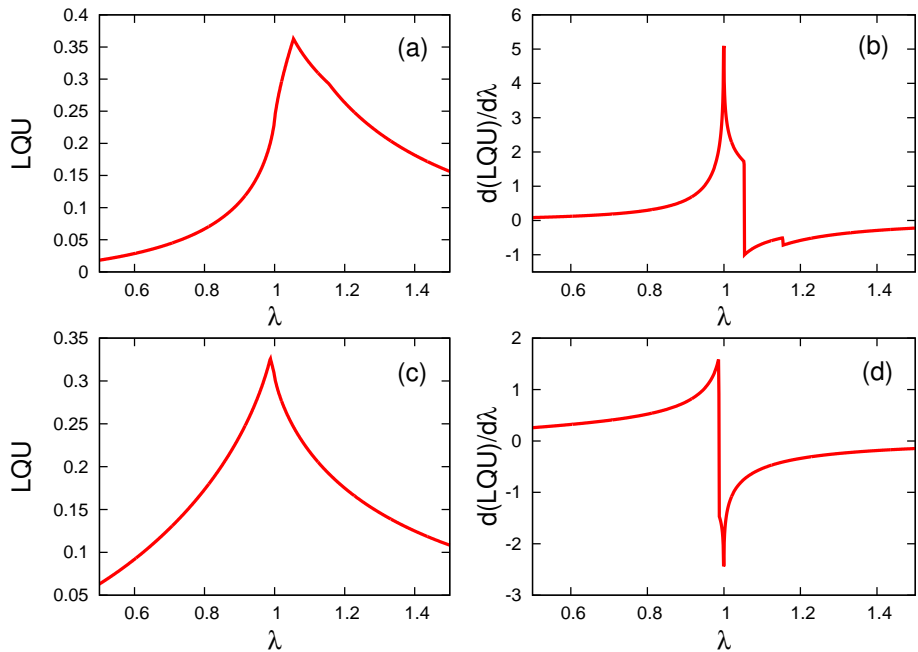}
\caption{Two-spin local quantum uncertainty for $\gamma=0.5$ (\textbf{a}) and $\gamma=1$ (\textbf{c}), along with its first derivative (with respect to $\lambda$) for $\gamma=0.5$ (\textbf{b}) and $\gamma=1$ (\textbf{d}), as a function of $\lambda$.}
\label{fig5}
\end{center}
\end{figure}

\section{Conclusions}

In this work, we have presented a review of the behavior of correlations in quantum spin chains and their relation to the phenomena of factorization and QPTs. We have provided an overview of the significant papers in the field, summarizing their main results. We have also included an extensive survey of our contributions to related problems.

In particular, we have firstly studied the thermal quantum and total correlations in the XY chain. There are four different measures considered in our investigation, introduced in Section 3, namely MIN, WYSIM, OMQC and concurrence. We have analyzed the behavior of these correlations near the CP of the QPT and seen that all four measures are able to detect the presence of the QPT at zero temperature. As we increase the temperature, measures start to lose their non-trivial behavior near the CP. Therefore, divergent derivatives are replaced with extrema points. However, looking at the location of these extrema, it is possible to determine an estimated CP at finite temperatures. In fact, we have used this method to compare the accuracy of the estimation performance of the considered measures at finite temperatures. We have seen that the performance of the measures in correctly estimating the CP of the QPTs is strongly dependent on the Hamiltonian parameters. For example, while OMQC has good performance in estimating the CP at finite temperature for $\gamma=0.5$, its accuracy significantly decreases for $\gamma=0.001$. Similar results have also been obtained for second nearest neighbors, but since they do not provide any new information about the system, we have not included them in our review. We have also explored the long-range behavior of the correlations and seen that in the ordered (ferromagnetic) phase, they remain non-zero for very large spin separations.

Second, we analyze the behavior of the coherence measure WYSI in the XY chain. We have seen that even if we look at the $\sigma_x$ coherence for a single spin, we can identify the QPT by the divergence in the first derivative. Furthermore, the experimentally-friendly simplified version of the coherence measure can also detect the QPT, which is an important result for possible experimental applications. On the other hand, both of them fail to recognize the factorized ground state. Moving on to the two-spin $\sigma_x$ local coherence, we again observe that the location of the CP can be spotlighted by the divergence in the measure. However, even though the simplified measure again does not feel the presence of the FP, LQC is able to pinpoint the position of the FP. Therefore, we have concluded that the discontinuous behavior of a correlation measure at the FP stems from the non-trivial behavior of the elements of the square root of the density matrix. It is important to note that using different observables from $\sigma_x$ does not provide any additional insight about the system under consideration. We have then turned our attention to the LQC, which is the optimized version of the LQC over all sets of observables. The divergence of the derivative of LQU is again present at the CP pinpointing the QPT. However, we see sharp extrema points, which do not correspond to any kind of non-triviality of the XY model. A detailed analysis of these points revealed that the extrema points are in fact caused by the change in the optimizing observable in the definition of LQC. Thus, it is important to be careful about the origin of non-analyticities showing-up in the behavior of quantum correlations before reaching conclusions about criticality and factorization.

\acknowledgments{Acknowledgments}

F.F.F. acknowledges the support from the National Institute for Science and Technology of Quantum Information (INCT-IQ) under grant 2008/57856-6, the National Counsel of Technological and Scientific Development (CNPq) under grant 474592/2013-8, and the São Paulo Research Foundation (FAPESP) under grant 2012/50464-0. G. K. is grateful to FAPESP for the support under grant numbers 2012/18558-5 and 2014/20941-7.

\newpage
\authorcontributions{Author Contributions}

All authors contributed equally to this paper.

\conflictofinterests{Conflicts of Interest}

The authors declare no conflict of interest.

%=================================================================
% References: Variant A
%=================================================================
% Back Matter (References and Notes)
%----------------------------------------------------------
% Style and layout of the references
\bibliographystyle{mdpi}
\makeatletter
\renewcommand\@biblabel[1]{#1. }
\makeatother

%=================================================================
% References:  Variant B
%=================================================================
% Use the following option to include external BibTeX files:
%\bibliography{lite}
%\bibliographystyle{mdpi}

%%%%%%%%%%%%%%%%%%%%%%%%%%%%%%%%%%%%%%%%%%

%\abbreviations{Abbreviations/Nomenclature}
%
%Main text.

%%%%%%%%%%%%%%%%%%%%%%%%%%%%%%%%%%%%%%%%%%

%\appendix
%\section{Appendix Title}
%
%Main text.

\end{document}